\documentclass[aps,pre,amsmath,amsfonts,floatfix,superscriptaddress,nofootinbib]{revtex4-2}
\usepackage{amsthm,amsmath}
\usepackage{graphicx}

\usepackage{hyperref}
\usepackage[utf8]{inputenc} 

\begin{document}



\title{adabmDCA: Adaptive Boltzmann machine learning for biological sequences}


\author{Anna Paola Muntoni \footnote{anna.muntoni@polito.it}}
\affiliation{
	Statistical Inference and Biological Modeling Group, Italian Institute for Genomic Medicine, 
	IRCCS Candiolo, SP-142, I-10060 Candiolo (TO), Italy
}

\author{Andrea Pagnani \footnote{andrea.pagnani@polito.it}}
\affiliation{
	Department of Applied Science and Technology (DISAT), Politecnico di Torino, 
	Corso Duca degli Abruzzi 24, I-10129 Torino, Italy
}
\affiliation{
	Statistical Inference and Biological Modeling Group, Italian Institute for Genomic Medicine, 
	IRCCS Candiolo, SP-142, I-10060 Candiolo (TO), Italy
}
\affiliation{Istituto Nazionale Fisica Nucleare (INFN), Sezione di Torino, Via Giuria 1, I-10125 Torino, Italy}

\author{Martin Weigt \footnote{martin.weigt@upmc.fr}}
\affiliation{
	Sorbonne Universit\'e, CNRS, Institut de Biologie Paris Seine, Biologie
	Computationnelle et Quantitative LCQB, F-75005 Paris, France
}

\author{Francesco Zamponi \footnote{francesco.zamponi@ens.fr}}
\affiliation{
	Laboratoire de Physique de l'Ecole Normale Sup\'erieure, ENS,
	Universit\'e PSL, CNRS, Sorbonne Universit\'e, Universit\'e de Paris,
	F-75005 Paris, France 
}

\begin{abstract} 

Boltzmann machines are energy-based models that have been shown to
provide an accurate statistical description of domains of evolutionary-related protein and RNA families. They are parametrized in terms of
local biases accounting for residue conservation, and pairwise terms
to model epistatic coevolution between residues. From the model
parameters, it is possible to extract an accurate prediction of the
three-dimensional contact map of the target domain. More recently, the
accuracy of these models has been also assessed in terms of
their ability in predicting mutational effects and generating \textit{in
silico} functional sequences.

Our adaptive implementation of Boltzmann machine learning,
\texttt{adabmDCA}, can be generally applied to both protein and RNA
families and accomplishes several learning set-ups, depending on the
complexity of the input data and on the user requirements. The code is
fully available at \url{https://github.com/anna-pa-m/adabmDCA}. As an
example, we have performed the learning of three Boltzmann machines
modeling the Kunitz and Beta-lactamase2 protein domains and
TPP-riboswitch RNA domain.

The models learned by \texttt{adabmDCA} are comparable to those
obtained by state-of-the-art techniques for this task, in terms of the
quality of the inferred contact map as well as of the synthetically
generated sequences. In addition, the code implements both equilibrium and out-of-equilibrium learning, which
allows for an accurate and lossless training when the equilibrium one
is prohibitive in terms of computational time,
and allows for pruning irrelevant parameters using an information-based criterion.

\end{abstract}


%




\maketitle
\section*{Background}

\subsection*{Protein and RNA sequence modelling}

In recent years, the number of available protein and RNA sequences
has shown an impressive growth thanks to the development of high-throughput
sequencing techniques.  As a consequence, databases like Pfam
\cite{mistry_pfam_2021} and Rfam \cite{kalvari_rfam_2021}, where such
biological sequences are annotated and classified according to
evolutionary relationships, play a dominant role in modeling this
enormous amount of data.

In spite of this constant increase of sequence data, the tertiary
structure of the corresponding domains is nowadays not experimentally
accessible for the majority of the known protein and RNA sequences. To
compensate for this experimental gap, \textit{in silico} protein and RNA
domains modeling has shown an incredible predictive power in
determining their structure~\cite{jumper2021highly,baek2021accurate}. An interesting way to achieve this
goal is to perform data-driven modeling to reproduce some relevant
statistical properties of the data (observables).
 
The so-called Direct Coupling Analysis (DCA)
\cite{weigt2009,morcos2011direct} turns out to be particularly successful in
using available homologous sequence data to infer structural
determinants of the underlying protein or RNA domains
\cite{cocco_inverse_2018}. In a nutshell, the DCA inference strategy
provides a simple and informative interpretation of the inferred set
of model parameters in terms of remarkably accurate contact map
prediction. Among the different DCA implementations, Boltzmann machine
learning~\cite{sutto_residue_2015,haldane_structural_2016-1,figliuzzi_how_2018,barrat-charlaix_sparse_2021} turns out to be one of the most
efficient in terms of: (i) the accuracy of structural predictions in its
direct use, or as input of more complex deep learning supervised
strategies~\cite{RaptorX2019,DeepMetaPsicov2019,AlphaFold2020,TrRosetta2020};
(ii) the effectiveness to generate artificially-designed sequences that fold
similarly to their natural counterparts~\cite{russ_evolution-based_2020}; (iii) the ability to predict mutational
effects~\cite{figliuzzi2016, hopf2017mutation}. However, the quality
of the Boltzmann machine deeply depends on the quality of the learning,
which is intrinsically linked to the way the model observables are
computed within the training, usually employing a Monte Carlo Markov
Chain (MCMC).

Here, we present \texttt{adabmDCA}, an adaptive Boltzmann machine
learning computational strategy that, taking as input a multiple
sequence alignment of a target protein or RNA domain, infers
efficiently an accurate statistical model of the sequence data with
the twofold aim of (i) providing an accurate contact map prediction of
the target domain, and (ii) generating artificial sequences that are
statistically close to indistinguishable (and thus {\em bona fide} biologically
functional) from the natural. The code implements the so-called
Boltzmann machine learning algorithm~\cite{ackley1985learning} by performing a gradient ascent
of the \textit{a posteriori} probability of the model given the input data. At variance with other implementations, \texttt{adabmDCA} copes with both protein and RNA data, and encompasses the ability of pruning redundant parameters (as described in~\cite{barrat-charlaix_sparse_2021}). Finally, it provides a standard \textit{equilibrium} learning and also a \textit{non-equilibrium} learning, based on the contrastive 
divergence technique~\cite{hinton2002training},
which is more suitable for structured data.

\subsection*{An introduction to Boltzmann  learning of biological models}

The input of Boltzmann learning is a multiple sequence alignment (MSA)
of a protein or a RNA family. The MSA contains $M$ aligned sequences
of length $L$ of the same domain family, characterized by a certain structure
and/or function.  The key idea behind DCA is that
evolutionarily related sequences show, together with a strong
conservation signal on key residues (for instance in presence of
active sites), large correlations among pairs of residues due to
(mostly) structural constraints. If a random mutation affects
the residue present on a certain site $i$, a compensatory mutation may be
needed to appear in sites $j$ in contact with residue $i$ in order to
preserve the overall structure of the domain. However, correlations
alone provide poor results in terms of contact prediction (due to the
presence of spurious correlations among sites not in direct contact),
and therefore more sophisticated techniques such as DCA have been used
to reliably extract the underlying coevolutionary signal.

In the following, we assume that each natural sequence
$\boldsymbol{s}=(s_1,...,s_L)$ belonging to a protein or RNA family is an
independent and identically distributed (i.i.d.) sample of an unknown
distribution (correlations due to phylogeny will be addressed through a standard re-weighting scheme explained below). Each residue $s_{i}$ takes value from an alphabet of
$q$ symbols representing all possible amino-acids or nucleic acids plus
a `-' gap symbol ($q = 21$ in the case of protein sequences or $q = 5$ for
RNA).  Given a MSA, we aim at finding a probability measure over
sequences of length $L$ that is able to accurately reproduce a set of
chosen observables computed from the data. In particular, Boltzmann
learning aims at reproducing all (or a subset of) the one-site and
two-site empirical frequency counts. By applying the maximum entropy
principle ~\cite{jaynes_information_1957}, the least-constrained distribution that characterizes our
data is a Boltzmann distribution:
\begin{equation}
P\left(\boldsymbol{s}\right|\mathbf{J}, \mathbf{h}) = \frac{1}{Z\left(
\mathbf{J}, \mathbf{h} \right)} \exp\left(\sum_{i} h_i\left(s_i\right)
+ \sum_{i<j} J_{i,j}\left(s_{i}, s_{j}\right) \right)\,,
\label{eq:Boltzmann}
\end{equation}
where $Z$ is the normalization factor (the so called partition
function in statistical-physics terminology) ensuring the proper
normalization of the distribution, and $\mathbf{J}$ and $\mathbf{h}$
are the set of Lagrange multipliers assuring the fit of the moments,
or, from a statistical physics point-of-view, the coupling matrices
and the fields associated with the Potts model in
Eq.~\ref{eq:Boltzmann}. This probability quantifies the likelihood that
a candidate sequence $\mathbf{s}$ belongs to the protein/RNA family
characterized by the set of parameters $\mathbf{J},\mathbf{h}$.
However, DCA aims at solving the related \textit{inverse} problem:
given a family's MSA, how to determine the set of
unknown parameters?

From a Bayesian perspective, the inverse problem can be mapped into
the problem of finding the set of couplings $\mathbf{J}$ and fields
$\mathbf{h}$ which maximize the posterior probability of
the unknown parameters given the set of observed configurations of
natural sequences
\begin{equation}
P\left(\mathbf{J},\mathbf{h}|\left\{ \boldsymbol{s}^{\mu}:\mu=1,\ldots,M\right\} \right)=\frac{P\left(\left\{ \boldsymbol{s}^{\mu}\right\} |\mathbf{J},\mathbf{h}\right)P\left(\mathbf{J},\mathbf{h}\right)}{P\left(\left\{ \boldsymbol{s}^{\mu}\right\} \right)}\,,
\end{equation}
where $P\left(\left\{ \boldsymbol{s}^{\mu}\right\} \right)$ is the
evidence (since it is independent of the unknown parameters, it will be
neglected in the following), $P\left(\left\{
\boldsymbol{s}^{\mu}\right\} |\mathbf{J},\mathbf{h}\right)$ is the
likelihood function, which describes how probable it is to randomly draw the sequences
in the MSA from the distribution parametrized by
$\left(\mathbf{J},\mathbf{h}\right)$; and $P\left( \mathbf{J},
\mathbf{h} \right)$ is the prior probability distribution over the space of
parameters. Recalling that sequences are i.i.d., the likelihood
factorizes over sequences:
\begin{eqnarray}
P\left(\left\{ \boldsymbol{s}^{\mu}\right\} |\mathbf{J},\mathbf{h}\right)	& = &\prod_{\mu=1}^{M}P\left(\boldsymbol{s}^{\mu}|\mathbf{J},\mathbf{h}\right)  \\
& = & Z^{-M}\left(\mathbf{J},\mathbf{h}\right)\prod_{\mu}e^{\sum_{i=1}^{L}h_{i}\left(s_{i}^{\mu}\right)+\sum_{i<j}J_{ij}\left(s_{i}^{\mu},s_{j}^{\mu}\right)}\,.
\end{eqnarray}
The prior $P\left(\mathbf{J},\mathbf{h}\right)$ provides additional
information on the unknown parameters and it is exploited to avoid
over-fitting. Indeed, very often a fully connected model (i.e. a model
for which all couplings $\mathbf{J}$ are in principle different from
0) produces an over-parametrization of the unknown distribution as
signaled by a large amount of noisy and negligible coupling parameters~\cite{barrat-charlaix_sparse_2021}. A practical way to control this
behavior is to impose a sparsity prior over the coupling matrices:
the two most used priors are the so called $\ell_{1}$ and $\ell_{2}$
regularizations, which force the inferred couplings to minimize the
associated $\ell_{1}$ and $\ell_{2}$ norms multiplied by a tunable
parameter $\lambda$ that sets the regularization strength. A
complementary approach consists in setting \textit{a priori} a
probable topology suggested by the mutual information between all
pairs of residues~\cite{gao_correlation-compressed_2018}. Here, as
discussed in the following section, we will follow an
information-based decimation protocol originally proposed in~\cite{barrat-charlaix_sparse_2021}.

To set the stage, we first start by discussing the case, in which no prior information is
considered. The maximization of the posterior distribution then turns out to be equivalent to the maximization of the likelihood function, or, equivalently, to the
log-likelihood:
\begin{equation}
\mathcal{L}\left(\left\{ \boldsymbol{s}^{\mu}\right\} |\mathbf{J},\mathbf{h}\right)=\frac{1}{M}\sum_{\mu}\left[\sum_{i=1}^{L}h_{i}\left(s_{i}^{\mu}\right)+\sum_{i<j}J_{ij}\left(s_{i}^{\mu},s_{j}^{\mu}\right)\right]-\log Z\left(\mathbf{J},\mathbf{h}\right)\,.
\end{equation}
It is easy to prove that the log-likelihood is a globally convex
function of the unknown parameters, hence a simple gradient ascent
strategy is in principle able to find the optimal set of parameters. More precisely, starting from any initial guess for the parameters $\left\{
\mathbf{J}^{t=0},\mathbf{h}^{t=0}\right\}$, one can set up the
following update scheme:
\begin{eqnarray}
\label{eq:upd1}
h_{i}^{t+1}\left(a\right)	& \leftarrow &	h_{i}^{t}\left(a\right)+ \eta_{h} \frac{\partial\mathcal{L}\left( \{\boldsymbol{s}^{\mu}\} | \mathbf{J}^{t}, \mathbf{h}^{t}\right)}{\partial h_{i}\left(a\right)} \ , \\
\label{eq:upd2}
J_{ij}^{t+1}\left(a,b\right) & \leftarrow &	J_{ij}^{t}\left(a,b\right)+ \eta_{J} \frac{\partial\mathcal{L}\left( \{\boldsymbol{s}^{\mu}\} | \mathbf{J}^{t}, \mathbf{h}^{t}\right)}{\partial J_{ij}\left(a,b\right)} \ ,
\end{eqnarray}
until a fixed point is reached. Here, $\eta_{h}$ and $\eta_{J}$
are the learning rates associated with the fields $\mathbf{h}$ and the
coupling parameters $\mathbf{J}$ respectively. A simple computation
shows that the gradient terms involve averages of simple observables
over the Boltzmann measure Eq.~\ref{eq:Boltzmann} with parameters at
iteration time $t$:
\begin{eqnarray}
\frac{\partial\mathcal{L}\left( \{\boldsymbol{s}^{\mu}\} | \mathbf{J}^{t}, \mathbf{h}^{t}\right)}{\partial h_{i}\left(a\right)} & = &	f_{i}\left(a\right) - p_{i}^{\left(t\right)}\left(a\right) \label{eq:likelihood-gradient1} \ , \\
\frac{\partial\mathcal{L}\left( \{\boldsymbol{s}^{\mu}\} |
  \mathbf{J}^{t}, \mathbf{h}^{t}\right)}{\partial
  J_{ij}\left(a,b\right)}	& = & f_{ij}\left(a,b\right) -
p^{\left(t\right)}_{ij}\left(a,b\right) \,.
\label{eq:likelihood-gradient2}
\end{eqnarray}
The stationary point is reached when the left hand sides of the
equations are zero, i.e. when the single and double residue empirical
frequency counts (i.e. the terms $f_i(a)$ and $f_{ij}(a,b)$ resp.)
match the one- and two-site marginals $p_i(a)$
and $p_{ij}(b)$ of the model $P$. A formal definition of these quantities will be given in the next sections. Unfortunately, in spite of the relatively simple
structure of the model, we are not able to exactly compute the
marginal probability distributions. A practical way to overcome this
limitation is to estimate them at each step $t$ of the iteration by
using a MCMC algorithm as explained in the following section.

\section*{Implementation}

\subsection*{Input data and pre-processing}
\label{sec:input}

\texttt{adabmDCA} takes as input a MSA in FASTA
format of the target protein or RNA family. To reduce the effect of
phylogenetic correlations, we re-weight the statistical significance
of each sequence, penalizing highly similar sequences in the MSA, as originally presented in~\cite{morcos_direct-coupling_2011}. In practice, with each of the $M$
sequences of the MSA we associate a statistical weight $w^\mu$
($\mu \in 1,\dots,M$) equal to the inverse number of sequences having at
least 80\% of identical residues with sequence $\mu$ (including sequence $\mu$ itself).

To deal with unobserved (pairs of) symbols in one (or two) column(s) of the MSA, we add a small pseudo-count $\alpha$ to
the empirical frequency counts. This prevents the emergence of
infinitely large parameters (in absolute value) associated with vanishing empirical frequencies. Finally, the one- and two-site frequencies are given by:
\begin{eqnarray}
f_{i}\left(a\right) & = & \left(1-\alpha\right) f^{\rm data}_{i}\left(a\right) + \frac{\alpha}{q} \ , \\
f_{ij}\left(a,b\right) & = & \left(1-\alpha\right) f^{\rm data}_{ij}\left(a,b\right) + \frac{\alpha}{q^{2}}\,,
\end{eqnarray}
where $f^{\rm data}_{i}$ and $f^{\rm data}_{ij}$ are computed from
the MSA as:
\begin{eqnarray}
\label{eq:emp_freq1}
f^{\rm data}_{i}\left(a\right) & = & \frac{1}{M_{\rm eff}}\sum_{\mu} w^{\mu} \delta_{s_{i}^{\mu},a} \ , \\
\label{eq:emp_freq2}
f^{\rm data}_{ij}\left(a,b\right) & = & \frac{1}{M_{\rm eff}} \sum_{\mu} w^{\mu} \delta_{s_{i}^{\mu},a} \delta_{s_{j}^{\mu},b}\,,
\end{eqnarray}
with $M_{\rm eff} = \sum_\mu w^\mu$ being the effective number of weighted sequences.

\subsection*{Initialization}

In \texttt{adabmDCA}, it is possible to initialize the set of
parameters in three ways: (i) all couplings and fields can be
initially set to zero, (ii) they can take value from a given set of
parameters (from an input file), or (iii) they describe a profile model, i.e. an
independent-site Potts model where the first empirical moments are
perfectly matched by means of the fields
\begin{equation}
h_{i}^{\rm prof}\left(a\right) = \log\left[f_{i}\left(a\right) \right] + {\rm const}\,,
\end{equation} 
but all couplings are set to zero.  Empirically, it turns out
that choice (iii) is the one that shows the fastest convergence
of the algorithm. We also allow for the other two types of initializations as they can be convenient in some cases.

\subsection*{Adaptive Monte Carlo Markov Chain} 
\label{sec:adaMC}
The Boltzmann learning algorithm consists of a series of training
epochs. At each epoch $t$, we estimate numerically the marginal
probability distributions of the model $p^{(t)}_i(a)$ and
$p^{(t)}_{ij}(a,b)$ using a MCMC strategy. More precisely, we use
$N_{s}$ independent Markov chains, each of which samples $N_{c}$ configurations. The results presented in this work are
obtained using a Metropolis-Hasting~\cite{metropolis_equation_1953, hastings_monte_1970} update scheme, but the code also allows one to opt
for a Gibbs sampling strategy~\cite{geman_stochastic_1984}. At the end
of each epoch $t$, we update the model parameters according to
Eq.~\ref{eq:likelihood-gradient2} by estimating the
$p^{(t)}_i(a), p^{(t)}_{ij}(a,b)$ according to the following relation:
\begin{eqnarray}
p_{i}^{\left(t\right)}\left(a\right) & = & \frac{1}{N_{s}N_{c} }\sum_{\mu=1}^{N_{s}N_{c}} \delta_{s_{i}^{\mu}\left(t\right), a} \ , \\
p_{ij}^{\left(t\right)}\left(a,b\right) & = & \frac{1}{N_{s}N_{c}} \sum_{\mu=1}^{N_{s}N_{c}} \delta_{s_{i}^{\mu}\left(t\right),a}\delta_{s_{j}^{\mu}\left(t\right),b}\,.
\end{eqnarray}

\texttt{adabmDCA} allows one to use either \textit{persistent} chains,
i.e. chains initialized only at the first epoch, or \textit{transient}
chains where each independent chain is initialized at each epoch. We consider two
types of chain initialization: (i) by extracting sequences uniformly
at random, (ii) by randomly picking natural sequences from the MSA, proportionally to their weights $\boldsymbol{w}$.

By default, \texttt{adabmDCA} uses \textit{transient} chains
initialized to uniformly extracted random sequences, but different
options can be set. In particular, we found that the
\textit{persistent} option seems to reduce the equilibration time as
one may expect that an equilibrium configuration extracted from the
model at time $t-1$ is a good candidate starting point for the same
chain at time $t$ provided that the value of the parameters at time
$t$ is not too different from that at $t-1$.

In order to achieve accurate learning, it is of utmost importance
to accurately estimate the gradient of the log-likelihood. From a
computational point of view, the bottleneck is the accurate estimation
of the one- and two-site marginals $p_i(a)$ and $p_{ij}(a,b)$. Two
main conditions dictate the quality of MCMC sampling: (i) an accurate
assessment of the stationary (i.e. equilibrium) regime of the chain,
and, (ii) a fair estimate of the mixing time \footnote{\texttt{adabmDCA} estimates the mixing time of the MCMC through the auto-correlation time of the sampled configurations}. To prevent the
occurrence of a poor sampling, \texttt{adabmDCA} allows for monitoring
and adjusting both the equilibration and sampling times of the Markov
Chain, $T_{\rm eq}$ and $T_{\rm wait}$, respectively (in Monte Carlo
sweeps units, one sweep being equal to $L$ Monte Carlo steps).

Let $\boldsymbol{s}^{i}_{n}$ be the configuration sampled by chain $i$
after $T_{\rm eq} + n T_{\rm wait}$ steps. We define three type of sequence identities or
\textit{overlaps}, i.e.
\begin{equation} 
O\left(\boldsymbol{s}^{i}_{n},
\boldsymbol{s}^{k}_{n}\right) :=
\sum_{j=1}^{L}\delta_{s^{i}_{n}\left(j\right),
  s^{k}_{n}\left(j\right)} \ ,
\end{equation}
aimed at quantifying how similar two
target configurations are:

\begin{itemize}
\item
  The \textit{external} overlap between configurations sampled by two
  different chains at the same sampling time $n$
\begin{equation*}
  Q^{\rm ext}\left(i,k,n\right) = O\left(\boldsymbol{s}^{i}_{n},
  \boldsymbol{s}^{k}_{n}\right)\, \mathrm{for}\, i\neq k \ .
\end{equation*}
\item
  A \textit{first-time internal} overlap measuring the similarity
  between two consecutively sampled configurations on the same chain:
\begin{equation*}
  Q^{\rm int1}\left(i,n,n+1\right) =
  O\left(\boldsymbol{s}^{i}_{n}, \boldsymbol{s}^{i}_{n+1}\right) \ .
\end{equation*}
\item
  A \textit{second-time internal} overlap measuring the
  distance between configuration sampled at time $n$ and $n+2$ on the
  same chain:
\begin{equation*}
  Q^{\rm int2}\left(i,n,n+2\right) =
  O\left(\boldsymbol{s}^{i}_{n},\boldsymbol{s}^{i}_{n+2}\right) \ .
\end{equation*}
\end{itemize}

At each iteration, we compute the expectation value $\mu_{\alpha}$ and
the standard error $\sigma_{\alpha}$ (where the averages are computed
with respect to different chains and over $n$) of $Q^{\alpha}$ for all
three types of overlap $\alpha \in \{\rm int1,\,\rm int2,\,\rm ext\}$.
We note that, if $T_{\rm eq}$ and $T_{\rm wait}$ were large enough, then subsequent samples of the same chain should have the same statistics of samples coming from distinct chains, and $\mu_{\rm ext} = \mu_{\rm int1} = \mu_{\rm int2}$ within statistical errors.
Therefore, we update $T_{\rm wait}$ as follows: 
\begin{itemize}
\item 
If $|\mu_{\rm ext} - \mu_{\rm int2}| >
5 \sqrt{\sigma^{2}_{\rm ext} + \sigma^{2}_{\rm int2}}$ we say that our
Monte Carlo chains are not sufficiently de-correlated and therefore we
increase $T_{\rm wait}$. 
\item
Conversely, if $|\mu_{\rm ext} - \mu_{\rm
  int1}| < 5 \sqrt{\sigma^{2}_{\rm ext} + \sigma^{2}_{\rm int1}}$ the
chains sufficiently de-correlate every $T_{\rm wait}$ steps and, as a
consequence, we can reduce $T_{\rm wait}$. 
\end{itemize}
This allows the chains to
be slightly correlated at time $T_{\rm wait}$ but ensures a good
de-correlation at time $2 T_{\rm wait}$, hence guaranteeing that the de-correlation time is in between $T_{\rm wait}$ and $2T_{\rm wait}$. To increase $T_{\rm wait}$,
we double it, while to reduce $T_{\rm wait}$,
\texttt{adabmDCA} computes the average between the current value of
the waiting time and the value of $T_{\rm wait}$ before the last
increasing step. This guarantees to keep the waiting time bounded in
the correct interval of values within the learning process.  Then,
whatever the outcome of this test, we set $T_{\rm eq} = 2 T_{\rm
  wait}$ assuming that $2 T_{\rm wait}$ steps suffice to get
equilibrated samples starting from the first configuration of the
chain. Note that when the starting
sample is picked uniformly at random, this criterion does not guarantee a perfect equilibration because the equilibration time might be in some cases larger than the de-correlation time, although this is expected to happen rarely;
conversely, for persistent chains,
this condition guarantees equilibration by construction, because in that case the chains do not need to be re-equilibrated at each iteration.

When $T_{\rm wait}$ and $T_{\rm eq}$ are such that $\mu_{ext} \sim
\mu_{int2}$, \texttt{adabmDCA} achieves a well-equilibrated sampling
and the Boltzmann machine is guaranteed to converge to a Potts model,
which not only precisely fits the one- and two-site frequencies,
but benefits of several additional properties elaborated in the Results section.  However,
depending on the properties of the data, several issues can arise: if
the true energy landscape is sufficiently rugged, Monte Carlo chains
may partially visit the feasible configurations returning a sampling
that strongly depend on the initialization of the chains. Similarly,
if the model parameters are abruptly adjusted, the dynamic may mimic a
low-temperature regime of a well-behaved landscape ending up to the
same sampling issue of the rugged energy landscape. In both cases, the
dynamics becomes non-ergodic, and therefore the computation of the
gradient may be inaccurate. For this reason a smooth update of the
parameters is encouraged and, in cases when this is not sufficient, we
found that, using persistent chains with fixed (but large) sampling
time, \texttt{adabmDCA} performs equally well. In this scenario, even
though the machine performs the sampling using slightly correlated
chains, the quality of the inferred model is often not affected. We
show an example in the Results section.

\subsection*{Convergence criterion and quality control}
\label{sec:convergence}

Given the global convexity of the problem as a function of the
parameters, the convergence of the algorithm can be safely assessed
when the gradients are numerically close to zero. A convenient proxy
for convergence is given by the difference between the empirical and
the model two-site connected correlations (or co-variances):
\begin{eqnarray}
  c_{ij}^{\rm model} \left(a,b\right) & = &
  p_{ij} \left(a,b\right) - p_{i}\left(a\right) p_{j}\left(b\right)
  \label{eq:data_conn_corr} \ , \\
  c_{ij}^{\rm emp} \left(a,b\right) & = &
  f_{ij} \left(a,b\right) - f_{i}\left(a\right) f_{j}\left(b\right)\,.
  \label{eq:emp_conn_corr}
\end{eqnarray}
The learning halts when the tolerance $\varepsilon_{c} = \arg
\max_{i,j,a,b} | \, c_{ij}^{\rm model}\left(a,b\right) - c_{ij}^{\rm
  emp}\left(a,b\right) \, | $ is $ \sim 10^{-2}$. Although this
quantity is not explicitly fitted during learning, it is a function of the one- and two-site frequencies in
Eqs. \ref{eq:likelihood-gradient1},\ref{eq:likelihood-gradient2} and vanishes at convergence. Empirically, it provides a good metric for estimating the quality of the
inferred model. At each iteration, we also measure the Pearson
correlation coefficient between the empirical and model covariances
defined in Eqs.~\ref{eq:data_conn_corr},\ref{eq:emp_conn_corr}, which measures a degree of
correlation between the two quantities
independently of the value of $\varepsilon_{c}$, i.e. of the spread of the
scatter plot of the connected covariance. 
Moreover, we display the fitting
error of the one- and two-site statistics computed as
$\varepsilon_{f} = \sum_{i,a} \frac{| f_{i}\left(a\right) - p_{i}
  \left(a\right) |}{Lq} $, $\varepsilon_{s} = \sum_{i,j,a,b} \frac{|
  f_{ij} \left(a, b \right) - p_{ij} \left(a,b \right) |}{L^2q^2}$ ;
these metrics indeed help in monitoring the training of the Boltzmann
machine.

Another interesting observable that can be used to assess the
generative power of the Boltzmann machine, is the three-site connected correlation 
\begin{equation*}
\begin{split}
  c_{ijk}(a,b,c):=&f_{ijk}(a,b,c)-f_{ij}(a,b)f_k(c)-f_{ik}(a,c)f_j(b) \\&-f_{jk}(b,c)f_i(a)+2f_i(a)f_j(b)f_k(c) \ ,
\end{split}
\end{equation*}
which is not fitted during the training but, as shown in~\cite{figliuzzi_how_2018,barrat-charlaix_sparse_2021}, provides an interesting measure of the generative capability of the model. \texttt{adabmDCA} does not compute all possible third order connected correlations because this would be computationally heavy. However, it is possible to specify a subset of indices $\left(i,j,k\right)$ and
$\left(a,b,c\right)$ whose corresponding measures are computed
during the iterations.

\subsection*{Gauge fixing}
\label{sec:gauge}
The number of unknown parameters $N_{p} =
\frac{L\left(L-1\right)}{2}q^{2}+Lq$ exceeds the number of independent
Eqs. \ref{eq:likelihood-gradient1},\ref{eq:likelihood-gradient2} (when
setting the partial derivatives to zero), due to the normalization
constraint on the one-site and two-site statistics,
$\sum_{a}f_{i}\left(a\right) = 1$, $\sum_{a,b} f_{ij}\left(a,b\right)
= 1$ and the marginalization condition over the two-site statistics,
$\sum_{a}f_{ij}\left(a,b\right) = f_{i}\left(b\right)$. As a
consequence, any \textit{gauge} transformation of the type
\begin{eqnarray}
J_{ij}\left(a,b\right) & \rightarrow & J_{ij}\left(a,b\right) + K_{ij}\left(a\right) + K_{ji}\left(b\right) \\
h_{i}\left(a\right) & \rightarrow & h_{i}\left(a\right) + g_{i} - \sum_{j\neq i}\left[K_{ij}\left(a\right) + K_{ji}\left(a\right)\right]
\end{eqnarray}
for arbitrary $g_{i}$ and $K_{i,j}\left(a\right)$, would keep
unchanged the Boltzmann distribution in Eq. \ref{eq:Boltzmann}. Among
the infinite number of possible \textit{gauge} transformations, the
one of most interest is the so-called \textit{zero-sum} gauge because
the couplings obtained by this re-parametrization minimize the
Frobenius norms associated with the coupling matrices. This
transformation is applied at the end of the Boltzmann machine learning
to facilitate the computation of the DCA scores.

Alternatively, one may fix the \textit{gauge} at the beginning of the
learning, by fixing a redundant subset of the parameters to an
arbitrary constant and then update the remaining parameters within the
learning. To select the redundant subset, for each couple
$\left(i,j\right)$, we seek the $2q-1$ pairs of colors that give the
weakest empirical connected correlations, computed as in
Eq. \ref{eq:emp_conn_corr}, and we set to zero the couplings
associated with these variables. These couplings are fixed to zero also during learning.

\subsection*{Pruning the parameters}
\label{sec:pru}
Although the gauge fixing removes the degree of variability of the
inferred parameters, due to the finite sample size of the MSA,
the trained model might still be over-fitted. Indeed, sequence lengths $L$ in typical MSA range in the interval $\sim 100 - 500$.  As
a consequence, the number of learned parameters is $\sim
10^{7}-10^{9}$, which likely exceeds the useful information encoded in
the data. A widely used strategy to limit over-fitting is to impose an
$\ell_{1}$ or $\ell_{2}$ regularization, i.e. a prior, either to both the set of
parameters or to the couplings only. In these cases, the update
Eqs.~\ref{eq:likelihood-gradient1}-\ref{eq:likelihood-gradient2} are replaced by the gradient of the log-posterior:
\begin{eqnarray}
\label{eq:l1}
\frac{\partial\mathrm{log}P \left(  \mathbf{J}^{t}, \mathbf{h}^{t} | \{\boldsymbol{s}^{\mu}\} \right)}{\partial h_{i}\left(a\right)} & = &	f_{i}\left(a\right) - p_{i}\left(a\right) -\lambda_{1} {\rm sign}  \left[h_{i}\left(a\right)\right] \ , \\
\label{eq:l1_2}
\frac{\partial\mathrm{log} P \left( \mathbf{J}^{t}, \mathbf{h}^{t} |  \{\boldsymbol{s}^{\mu}\} \right)}{\partial J_{ij}\left(a,b\right)}	& = & f_{i,j}\left(a,b\right) - p_{i,j}\left(a,b\right) - \lambda_{1} {\rm sign} \left[J_{i,j}\left(a,b\right)\right]\,, 
\end{eqnarray}
and 
\begin{eqnarray}
\label{eq:l2}
\frac{\partial\mathrm{log} P \left( \mathbf{J}^{t}, \mathbf{h}^{t} | \{\boldsymbol{s}^{\mu}\} \right)}{\partial h_{i}\left(a\right)} & = &	f_{i}\left(a\right) - p_{i}\left(a\right) -\lambda_{2} h_{i}\left(a\right) \ , \\
\label{eq:l2_2}
\frac{\partial\mathrm{log} P \left( \mathbf{J}^{t}, \mathbf{h}^{t}  |  \{\boldsymbol{s}^{\mu}\} \right)}{\partial J_{ij}\left(a,b\right)}	& = & f_{i,j}\left(a,b\right) - p_{i,j}\left(a,b\right) - \lambda_{2} J_{i,j}\left(a,b\right) \  , 
\end{eqnarray}
for the $\ell_{1}$ and $\ell_{2}$ priors respectively.

The main drawback of these procedures is that the regularization is
applied indistinctly to all parameters (relevant and irrelevant).
Alternatively, one may \textit{a priori} specifically prune (viz. set
to zero) a subset of the parameters observing that even though large
spurious correlations may arise from non topologically connected
sites, weak correlations are typically associated with small coupling
strengths. As explained in~\cite{gao_correlation-compressed_2018},
one can first determine a starting topology and then run the learning
procedure on it. To this end, \texttt{adabmDCA} provides two
distinct strategies. Indeed, the user can:
\begin{itemize}
\item provide as input a given topology (i.e. a set of predefined
  pairs of residues that will not be set to 0); \texttt{adabmDCA} then automatically
  eliminates all absent parameters before the learning;
\item iteratively remove negligible couplings up to a target
  sparsity as explained in~\cite{barrat-charlaix_sparse_2021}. To
  determine whether a coupling matrix (or element) is negligible, we
  compute the symmetric Kullback-Leibler divergence between the model
  at the current time-step and the same model without that coupling
  matrix (or element). The latter is used to score the parameters and
  set to
  zero those with the smallest score. The parameter is set to zero \textit{element-wise} if we
  remove negligible couplings drawn on different matrices or
  \textit{block-wise} if we remove an entire $\mathbf{J}_{ij}$
  matrix. We refer to~\cite{barrat-charlaix_sparse_2021} for details
  of the element-wise decimation.
  
\end{itemize}

\subsection*{Adaptive learning rate}

\label{seq:ada_learn}
The learning rates $\eta_{J}$ and $\eta_{h}$ associated with the
update of the fields and couplings, respectively, are set by default to
a small and constant value, typically $0.05$ for proteins and $0.01$
for RNA families. Alternatively, several adaptive learning rates can
be used to train the Boltzmann machine: \textit{adagrad}~\cite{duchi_adaptive_nodate},
\textit{search-then-converge}~\cite{darken_note_1990}, a modified
\textit{quasi-Newton} method~\cite{ferguson_translating_2013, haldane_structural_2016, haldane_mi3-gpu_2021} and \textit{FIRE}~\cite{bitzek_structural_2006}. Although using an adaptive learning
rate allows for a fast training of the parameters (as indicated by a
rapid increasing of the Pearson correlation coefficient between the
data and model covariances already in the first few iterations), the
possibly large learning rates push the value of the parameters to
large (absolute) values preventing a good equilibration of the machine
within the training, and often resulting in over-fitting.

\subsection*{Schematic workflow}

\begin{figure}[h]
    \centering
    \includegraphics[width=0.87\textwidth]{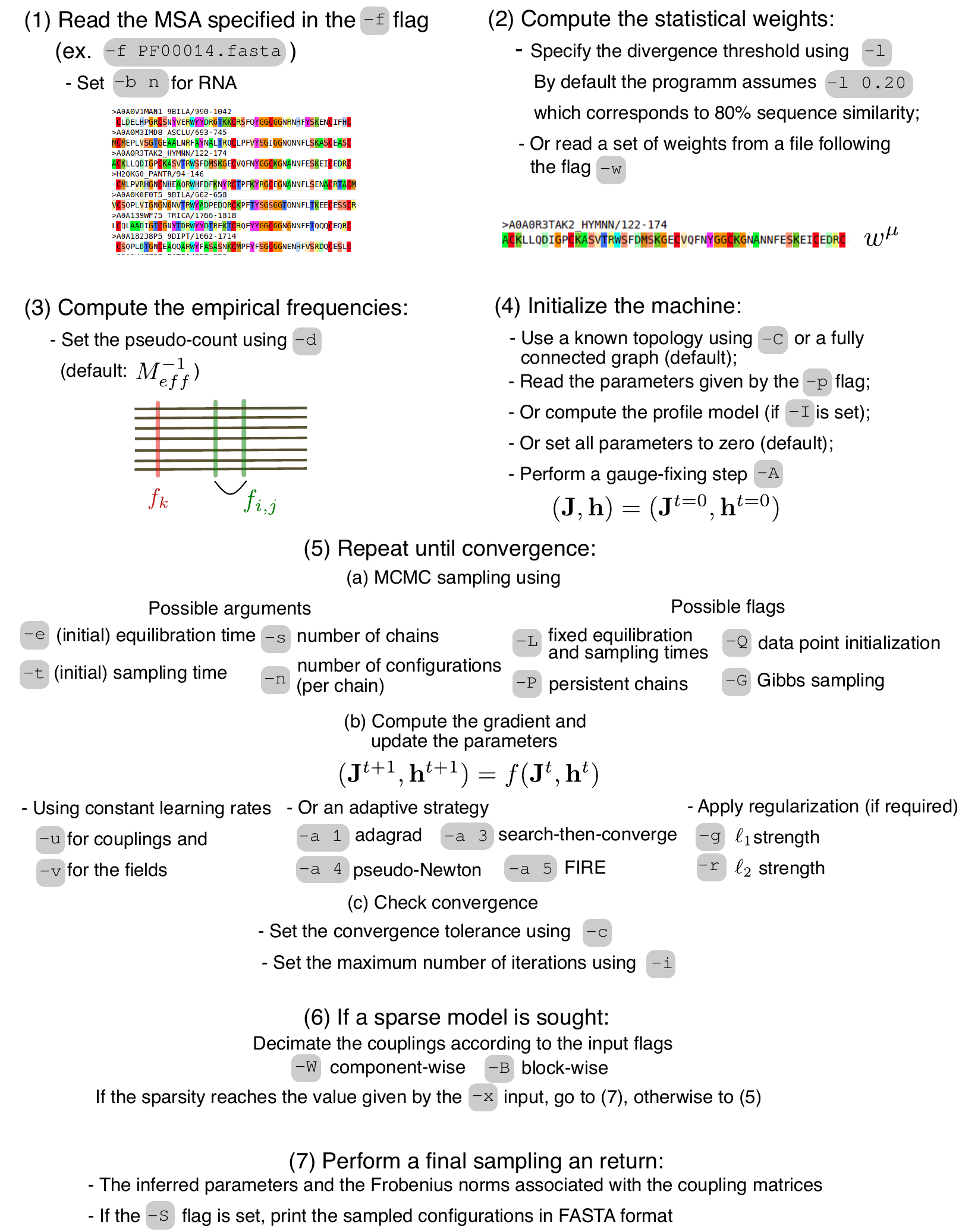}
    \caption{\label{fig:pseudo_code} \textbf{Workflow of \texttt{adabmDCA}.} We show a schematic road-map of the main \texttt{adabmDCA} workflow specifying the input flags associated with the possible tasks that can be executed by the algorithm. Lower case flags are usually followed by an argument whereas upper case flags appear alone.}
\end{figure}

To clarify the main \texttt{adabmDCA} road-map we plot in Fig. ~\ref{fig:pseudo_code} a schematic representation of the features performed by the algorithm (as well as the most important input flags):
\begin{itemize}
\item{\textbf{Reading the natural sequences.} The algorithm first reads a FASTA file containing the multiple sequence alignment of protein or RNA sequences.}
\item{\textbf{Re-weighting of the sequences.} \texttt{adabmDCA} either takes as input a file storing the statistical weights of the sequences or it applies the re-weighting scheme explained in Section \nameref{sec:input}. }
\item{\textbf{Computation of the observables.} Once the weights are computed, it is possible to evaluate and store the one-site and two-site frequencies appearing in the log-likelihood (or log-posterior) as in Eqs. \ref{eq:emp_freq1}}-\ref{eq:emp_freq2}. The pseudo-count $\alpha$ can be arbitrarily set or, by default, it takes the value of $M_{eff}^{-1}$.
\item{\textbf{Initialization of the machine.} By default, the machine assumes a fully connected model and the parameters are set to zero. Alternatively, a profile model can be chosen using a pre-defined flag or the machine can read an input set of parameters from a file. The gauge-fixing procedure, as explained in Section \nameref{sec:gauge} can be performed using a specific flag. Furthermore, in cases where the topology is known, \texttt{adabmDCA} can read from files the (possibly non-zero) couplings and the fields of the machine and set permanently to zero the remaining part.}
\item{\textbf{Update of the parameters until convergence.} At each epoch, \texttt{adabmDCA} performs a MCMC sampling as described in Section \nameref{sec:adaMC} to estimate the model statistics. All possible flags used to set up the MCMC sampling are shown in Fig. ~\ref{fig:pseudo_code}. By default, the equilibration and sampling times are adaptively tuned as described in Section \nameref{sec:adaMC}. Then, the parameters are updated accordingly to the gradient as in Eqs. ~\ref{eq:upd1}-\ref{eq:upd2} (or as in Eqs. ~\ref{eq:l1}-\ref{eq:l1_2} or ~\ref{eq:l2}-\ref{eq:l2_2}) depending on the presence (or absence) of the regularization terms. The learning rate is by default constant during the training but, if required by the user, several adaptive learning strategies are implemented (see Section \nameref{seq:ada_learn}). }
\item{\textbf{Decimation.} If required and if convergence is reached, \texttt{adabmDCA} performs a component-wise or block-wise pruning of the coupling matrices according to an information-based criterium (see Section \nameref{sec:pru} ). Then, the algorithm alternates the convergence step to the pruning step of the Boltzmann machine until a converged model having the required density is sought.}
\item{\textbf{Output of the results.} The algorithm performs a final sampling of the converged Boltzmann machine and prints in several files the couplings and fields of the model as well as the Frobenius norms, i.e. the Direct Coupling scores, associated with the $\mathbf{J}_{ij}$ matrices. If required by the user \texttt{adabmDCA} outputs the sampled configurations in FASTA format. }
\end{itemize}

 In the following, we report few examples to launch \texttt{adabmDCA} in some interesting cases, useful to reproduce the results of the following section:
 \begin{itemize}
 \item{\textit{Learning at equilibrium.} Let us train a Boltzmann machine for the sequences contained in \texttt{file.fasta} at equilibrium, starting from a profile model and requiring a tolerance of $10^{-2}$ for the two-site connected correlations, using our machine. The command line will read:
 \begin{equation}
 \texttt{./adabmDCA -f file.fasta -I -c 1e-2}
 \end{equation}
 }
 \item{\textit{Learning out-of-equilibrium.} To use persistent chains and avoid the tuning of the MCMC characteristic times, we will add:
 \begin{equation}
 \texttt{./adabmDCA -f file.fasta -I -c 1e-2 -L -P}
 \end{equation}
 }
 \item{\textit{Sampling.} Let us sample a given model stored in the file \texttt{p.dat} using $T_{eq} = 500$ and $T_{wait} = 250$. The command line reads:
 \begin{equation}
 \texttt{./adabmDCA -f file.fasta -p p.dat -i 0 -e 500 -t 250 -L -S}
 \end{equation}
 }
 \end{itemize}

\section*{Results}

We now discuss some examples of model learning via \texttt{adabmDCA} on protein and RNA families.

\subsection*{Learning at equilibrium: PF00014 and RF00059}

In this section we show the results obtained for: (i) the Kunitz
domain (PF00014 family from the Pfam database), (ii) the TPP
riboswitch (RF00059 from the Rfam database).  The PF00014 MSA is
initially pre-processed to remove from the MSA all proteins with more
than six consecutive gaps. This prevents a learning bias towards very
gapped configurations. Eventually, the total number of considered
sequences is $M = 13600$ for PF00014 and $M = 12593$ for RF00059,
which correspond to a re-weighted effective number of sequences of
$M_{\rm eff} = 4364$ and $M_{\rm eff} = 4920$ for PF00014 and RF00059
respectively.

\begin{figure}[ht]
    \centering
    \includegraphics[width=0.97\textwidth]{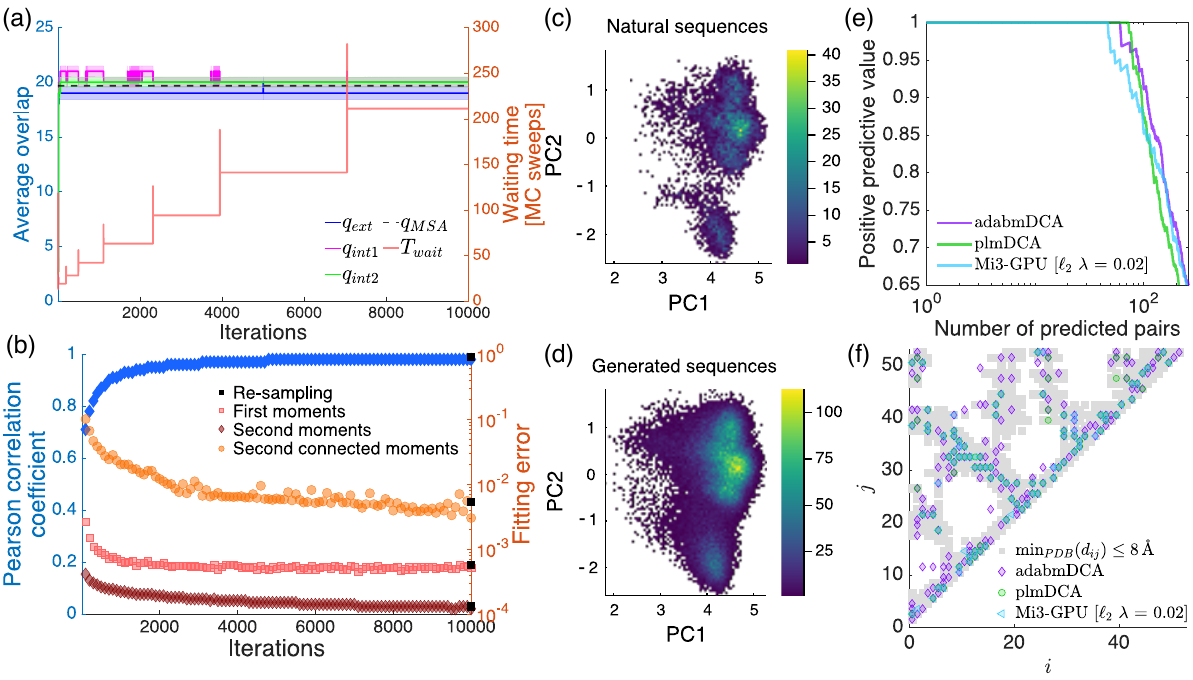}
    \caption{\label{fig:14eq} \textbf{Learning of PF00014 at equilibrium} We show in panel (a) the evolution over the iterations of the three overlaps used to monitor the quality of the sampling together with the waiting time $T_{\rm wait}$. In panel (b) we plot, for all iterations, the fitting errors (red, orange and brown markers) associated with the one-site, two-site (connected and non-connected) frequencies computed as defined by $\varepsilon_{f}$, $\varepsilon_{s}$ and $\varepsilon_{c}$ in Section~\nameref{sec:convergence}. Using a blue marker we show the Pearson correlation coefficient between the two-site connected frequencies of the natural sequences and of the configurations generated during training. In (c) we plot the projections of the natural sequences into the space of the first two principal components (PC1, PC2) of the covariance matrix associated with the natural sequences while in (d) we project the configurations obtained by the re-sampling of the converged model into PC1 and PC2 associated with the natural sequences. Panel (e) depicts the behavior of the positive predictive value (PPV) versus the number of non-trivial contact predictions, i.e. those associated with site indices $|i-j|>4$, for \texttt{adabmDCA}, \texttt{plmDCA}~\cite{ekeberg_improved_2013} and \texttt{Mi3-GPU}~\cite{haldane_mi3-gpu_2021}. In panel (f) we instead plot the contact maps used for the comparison in panel (e): gray blocks are associated with the ground-truth obtained by \texttt{Pfam-interactions}~\cite{edoardo_sarti_2020_4080947}, while the colored markers indicate whether the Frobenius norms computed using the parameters retrieved by the three methods are larger than 0.20.}
\end{figure}

\begin{figure}[bt]
    \centering
    \includegraphics[width=0.97\textwidth]{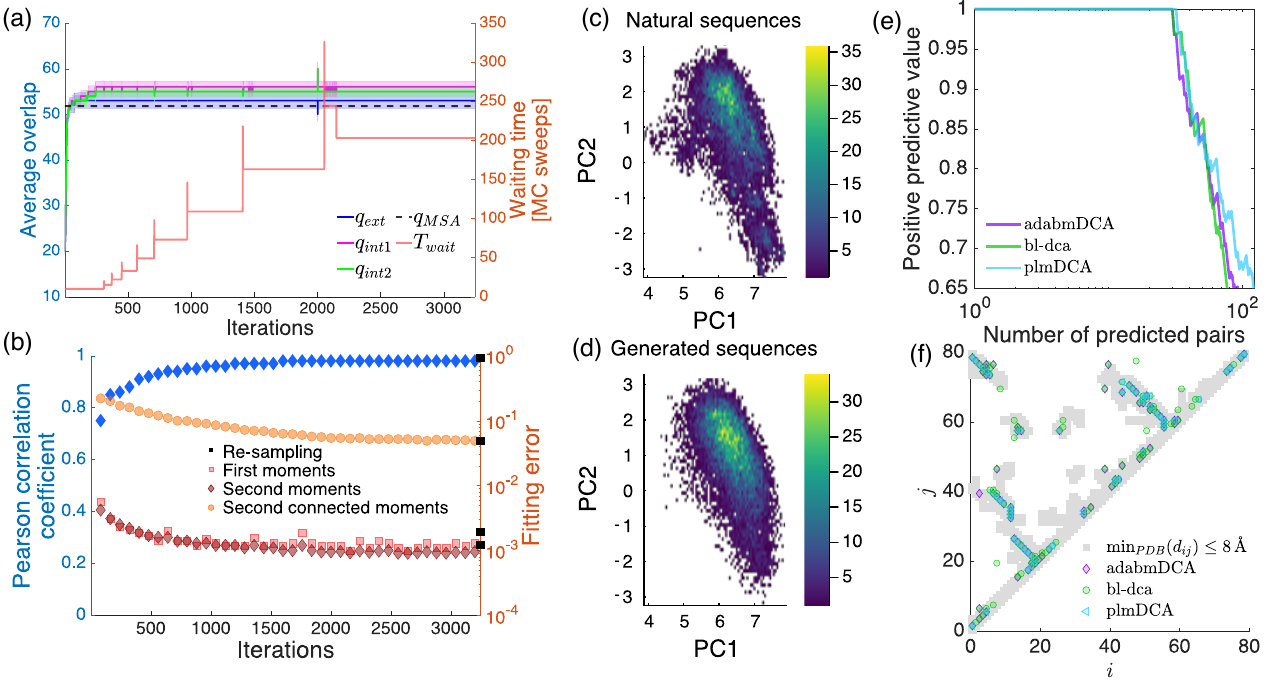}
    \caption{\label{fig:RF59}\textbf{Learning of RF00059 at equilibrium} Panel (a) shows the evolution over the iterations of $q_{ext}$, $q_{int1}$ and $q_{int2}$ used to tune the waiting time $T_{wait}$. In panel (b) we plot the fitting errors (red, orange and brown markers) $\varepsilon_{c}$, $\varepsilon_{f}$ and $\varepsilon_{s}$, and the Pearson correlation coefficient between the two-site connected statistics of the natural sequences and of the configurations sampled during training (blue markers). Panel (c) depicts the projections of the natural sequences into the space of the first two principal components (PC1, PC2) of the covariance matrix associated with the natural sequences; in (d) we show the projection the re-sampled configurations, obtained from the converged model, into PC1 and PC2 associated with the natural sequences. In panel (e) we show the behavior of the PPV versus the number of non-trivial contact predictions, i.e. those associated with site indices $|i-j|>4$, for \texttt{adabmDCA}, \texttt{plmDCA}~\cite{ekeberg_improved_2013} and \texttt{bl-dca}~\cite{cuturello_assessing_2020}. Panel (f) displays the contact maps used as ground truth (gray markers) for the TPP riboswitch and those obtained by the DCA scores larger than 0.20 associated with the three compared methods.}
\end{figure}

The Boltzmann machines are trained at equilibrium, i.e. the waiting
and equilibrium times are updated at each iteration according to the
test introduced in Section \nameref{sec:adaMC}. The behavior of the
average of the three overlaps $q_{\alpha}$ for $\alpha = \{\rm{int1},
\rm{int2}, \rm{ext}\}$ is shown in Figs.~\ref{fig:14eq}(a) and \ref{fig:RF59}(a) (left axis) together with the trend of the
waiting time $T_{\rm wait}$ (right axis). One can see that the
distribution of the mean for the three quantities show statistically
compatible values. Interestingly, starting from the beginning of the
training, their average value is very close to the mean overlap among
all pairs of natural sequences used within the learning, shown as
$q_{MSA}$ in the plot. The waiting time is typically increased at the
beginning of the training and it seems to stabilize at the final
iterations.

The quality of the Boltzmann machine is monitored during the
learning as shown in Fig.~\ref{fig:14eq}(b) and
Fig.~\ref{fig:RF59}(b) where we display the Pearson correlation
coefficient between the model and the empirical connected two-site
frequencies as computed in Eqs. \ref{eq:emp_conn_corr} (left axis) as
a blue line together with the mean error achieved in fitting the
one-site and (connected or non-connected) two-site frequencies (right
axis). At the final iteration we get a very accurate model as signaled
by the high value of the Pearson correlation coefficient and the small
values of the fitting errors, which are perfectly retrieved if one
samples the final models using a very long Monte Carlo Markov Chain
(black squared point), i.e. by imposing $T_{\rm eq} = 5000$ and $T_{\rm wait}
= 2500$. The generative power of the Boltzmann machines is
corroborated by comparing the Principal Component Analysis (PCA) of
the generated sequences with the natural sequences as shown in
Figs.~\ref{fig:14eq},\ref{fig:RF59}(c) and (d) respectively. The
sampled configurations in panel (d) are projected onto the first two
principal components of the natural sequences in panel (c). As
suggested by the spatial localization of the sequences and their
distribution, our converged models are able to generate sequences that
lie in the same non-trivial sub-space spanned by the natural
sequences.

Finally, we compare the predicted contact maps of the Kunitz domain
and of the TPP riboswitch with the following state-of-the-art
DCA-based algorithms: \texttt{plmDCA}~\cite{ekeberg_improved_2013}
and \texttt{Mi3-GPU}~\cite{haldane_mi3-gpu_2021} for PF00014 and
\texttt{bl-dca}~\cite{cuturello_assessing_2020} for RF00059, one pseudo-likelihood method and two
Boltzmann machine-based methods to infer Potts models for protein and
RNA sequences respectively.

In all cases, coupling parameters are first converted to zero-sum
gauge before computing the average product corrected Frobenius norms~\cite{ekeberg_improved_2013}. For PF00014, we consider as
ground-truth the atomic distances retrieved by
\texttt{Pfam-interactions}~\cite{edoardo_sarti_2020_4080947}, a method
which computes the minimum distance, for all possible pair of sites,
among all available crystal structures in the Protein Data Bank
(PDB). For RF00059 we perform an analogous analysis among the TPP
riboswitch known structures downloaded from the Protein Data Bank. In
Figs.~\ref{fig:14eq},~\ref{fig:RF59}(e) we plot the positive predictive value of the prediction of the non-trivial contacts, i.e. those residue
pairs $(i,j)$ having $|i-j| > 4$, for the three methods, and in
Figs.~\ref{fig:14eq},~\ref{fig:RF59}(f) we overlap our ground-truth
(in gray) and the most probable contact according to the
three-methods, i.e. the pairs with whom we associate a score larger
than 0.20. For \texttt{Mi3-GPU} we consider the model obtained
applying an $\ell_{2}$ regularization with strength parameter $\lambda
= 0.02$; the machine obtained for the $\ell_{1}$ regularization gives a dramatically worse results in terms of
contact predictions (not shown). Panel (e) suggests that the three
considered methods achieve comparable performances, as it is
equivalently represented in panel (f).

\subsection*{Learning out-of-equilibrium: PF13354}

\begin{figure}[ht]
    \centering
    \includegraphics[width=0.97\textwidth]{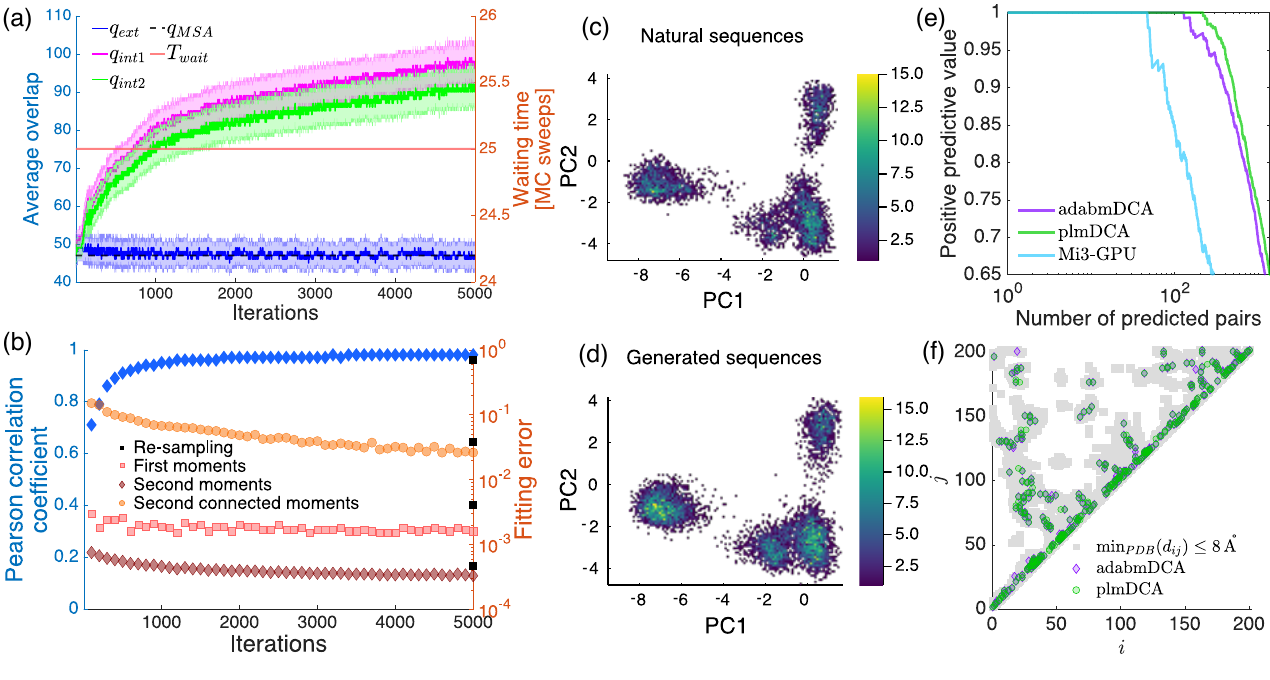}
    \caption{\label{fig:TEM1} \textbf{Learning of PF13354 out-of-equilibrium}. In panel (a) we show the three overlaps $q_{ext}$, $q_{int1}$ and $q_{int2}$ of the sampled configurations used to estimate the model statistics as a function of the iterations (left axis) and the waiting time $T_{\rm wait}$ between two consecutive samples (right axis). At difference with the learning at equilibrium, $T_{\rm wait}$ is here kept constant during the training and the configurations are correlated as suggested by the differences between the distributions of $q_{int1}$, $q_{int2}$ and $q_{ext}$. Panel (b) shows the plot of the quality metrics used to estimate the goodness of the training: in blue we show the Pearson correlation coefficient between the two-site connected frequencies of the natural sequences and of the evolving model as a function of the iterations (blue markers, left axis) and the fitting errors (red, orange and brown markers, right axis) computed as $\varepsilon_{c}$ for the two-site connected statistics and as $\varepsilon_{f,s}$ for the one-site and two-site non-connected statistics. In panels (c) and (d) we show the projections of the natural sequences and of the re-sampled sequences into the first two principal components of the natural sequences while in panel (e) we plot the positive predictive value curve associated with the contact map prediction (shown in panel (f)) for the Beta-lactamase2 domain.}
\end{figure}

In this section we show the results obtained for the Beta-lactamase2
domain. The multiple sequence alignment used within
the training is constructed as follows. Using the Hidden Markov Model
associated with the PF13354 family, we scanned the NCBI~\cite{NCBI}
database to obtain aligned sequences compatible with the model. We
then keep sequences that have less than 20\% of gaps and concurrently
those having less than 80\% redundancy (as a consequence $M_{\rm eff} \sim
M$ in this case). We also removed the sequence of the TEM-1 protein,
and all sequences very similar to it. This last step was necessary to
study deep mutational scanning data in~\cite{barrat-charlaix_sparse_2021} and we use here the same alignment
for sake of simplicity. Training a Boltzmann machine using 
well-equilibrated Monte Carlo chains is barely practical as the
waiting time necessary to produce uncorrelated samples is huge and
constantly increasing over the iterations (not shown). To solve this
issue, we resort to a \textit{persistent} sampling strategy, i.e. at
each new iteration the Monte Carlo chains are initialized at the last
configurations of the previous iteration, of $10^{3}$ chains, each one
sampling $10$ configurations, with fixed waiting time $T_{\rm wait} =
25$ and equilibrium time $T_{\rm eq} = 50$ sweeps. In
Fig.~\ref{fig:TEM1}(a) we display the overlap between independent
chains $q_{ext}$, which is similar to that of the MSA of the natural
sequences $q_{MSA}$, while $q_{int1}$ and $q_{int2}$ grow over the
iterations suggesting that the samples are highly correlated. The
fitting quality of the model is measured by using the Pearson
correlation coefficient (blue markers) and the fitting errors over the
one-site and two-site statistics (red, orange and brown markers) as shown in
Fig.~\ref{fig:TEM1}(b); these measures are compatible to those
obtained by a learning at equilibrium. To check the quality of the
learning out-of-equilibrium, we re-sample the converged model and test
the generative properties of the learned machine. The Pearson
correlation coefficient and the fitting errors of the converged model
are retrieved only if the configurations obtained by the re-sampling
step are sufficiently de-correlated: indeed, to obtain the
performances shown using black markers in Fig.~\ref{fig:TEM1}(b) one
has to set $T_{\rm wait} \sim 10^{5}$ which is the value of the waiting
time that guarantees $q_{int1} \sim q_{int2}$. The remarkable results
of the Beta-lactamase2 model are confirmed by the PCA analysis in
Fig.~\ref{fig:TEM1}(c) and (d) and by the contact prediction
depicted in panels (e) and (f). In this case, \texttt{adabmDCA}
achieves a reconstruction similar to that of \texttt{plmDCA} and
outperforms \texttt{Mi3-GPU} where the adaptive strategy to sample
statistically independent equilibrium configurations fails to produce
a result due to the too large auto-correlation time estimated.

This result suggests that although we are not able to achieve an
equilibrium sampling due to the large auto-correlation
time, yet the resulting model retains the generative properties of an
equilibrium-trained Boltzmann machine. Not only this result is
important from a practical point of view, as this allows for a
significant reduction of the computational time of the overall
process, but it also opens new research directions in the field of
out-of-equilibrium learning. We mention that if the procedure is
performed using randomly initialized chains, instead of persistent
chains, the quality of the converged model is achieved only setting a
waiting time similar to that used in the training, as if the model had
kept memory of the learning set-up. A similar behavior has been
observed and discussed more systematically in~\cite{decelle_equilibrium_2021} in the context of
learning Restricted Boltzmann machines.

\subsection*{Running time}
We discuss in this section the computation time of \texttt{adabmDCA}. The running time needed by \texttt{adabmDCA} is often larger than the ones shown by the other methods used here for comparison: our machine spent 22, 53, and 98 hours for learning a model for PF00014, RF00059, and PF13354, respectively, against one hour and the 75 hours required by \texttt{Mi3-GPU} for PF00014 and PF13354 (employing two TITAN RTX GPUs) and the two hours needed by \texttt{bl-dca} for RF00059. We stress that the current implementation exploits a single thread and its running times are compatible with those achieved by the Boltzmann machine in~\cite{figliuzzi_how_2018}. Moreover, the out-of-equilibrium learning allows for the training of an accurate machine, out-performing \texttt{Mi3-GPU} and spending a running time which is only slightly larger than that needed by \texttt{Mi3-GPU}, a highly optimized algorithm. \\
Fortunately, a multi-threads implementation of \texttt{adabmDCA} can be certainly attained by running in parallel the MCMC sampling, i.e. each thread could perform the simulations of a certain fraction of the MC chains, independently of the other threads. This direction will be considered in the future development of the algorithm.

\section*{Conclusions}
We developed a C/C++ implementation of Boltzmann machine learning for modeling RNA and protein sequence families, called \texttt{adabmDCA}. Together with a set of learning options that allows for a user-friendly control of the training strategy (including parameters initialization, regularization and decimation), it encompasses the possibility of adapting the Monte Carlo Markov Chain sampling ensuring an equilibrium training. In hard learning regimes, when the de-correlation time of the Monte Carlo chains appears to be large, the learning at equilibrium is intractable. In these cases, in \texttt{adabmDCA} it is possible to select a slightly out-of-equilibrium sampling whose behavior does not affect the quality of the learned model, as suggested by the results on the Beta-lactamase2 domain. Here, the performances of \texttt{adabmDCA} resemble those of \texttt{plmDCA} in predicting non-trivial physical contacts
and outperforms other Boltzmann machine-like implementations.
This promising achievement encourages new research perspectives in the field of non-equilibrium learning.

\section*{Availability and Requirements}
Project name: \texttt{adabmDCA} \\
Project home page: \url{https://github.com/anna-pa-m/adabmDCA} \\
Operating systems: Linux, Mac OS and Windows \\
Programming languages: C/C++ \\
Licence: MIT Licence \\
Any restriction to use by non-academics: No. 

\section*{List of Abbreviations}
DCA: Direct coupling analysis, PCA: Principal Components Analysis, MSA: multiple sequence alignment, PPV: positive predictive value, MCMC: Monte Carlo Markov Chain.

\section*{Declarations}

\section*{Ethics approval and consent to participate}
Not applicable.

\section*{Consent for publication}
Not applicable. 

\section*{Availability of data and materials}
The multiple sequence alignments analysed during the current study are available in the GitHub repository \url{https://github.com/anna-pa-m/adabmDCA}.

\section*{Competing interests}
The authors declare that they have no competing interests.

\section*{Funding}
This work was supported by a grant from the Simons Foundation (\#454955, Francesco Zamponi), and from the EU H2020 research and innovation program MSCA-RISE-2016 (grant 734439 InferNet). The funders have played no role in the design of the study and collection, analysis, and interpretation of data and in writing the manuscript.

\section*{Authors' contributions}
A.P.M, A.P., M.W. and F.Z designed the research; A.P.M. and F.Z. implemented the code; A.P.M., A.P. and F.Z. analyzed the data and wrote the paper. All authors have read and approved the final manuscript.

\section*{Acknowledgements}
We thank Beatriz Seoane and Aur\'elien Decelle for interesting
discussions on out-of-equilibrium learning strategies and for sharing the results of Ref.~\cite{decelle_equilibrium_2021} prior to publication.


\bibliographystyle{bmc-mathphys} 


\newcommand{\BMCxmlcomment}[1]{}

\BMCxmlcomment{
	
	<refgrp>
	
	<bibl id="B1">
	<title><p>Pfam: The protein families database in 2021</p></title>
	<aug>
	<au><snm>Mistry</snm><fnm>J</fnm></au>
	<au><snm>Chuguransky</snm><fnm>S</fnm></au>
	<au><snm>Williams</snm><fnm>L</fnm></au>
	<au><snm>Qureshi</snm><fnm>M</fnm></au>
	<au><snm>Salazar</snm><fnm>GA</fnm></au>
	<au><snm>Sonnhammer</snm><fnm>ELL</fnm></au>
	<au><snm>Tosatto</snm><fnm>SCE</fnm></au>
	<au><snm>Paladin</snm><fnm>L</fnm></au>
	<au><snm>Raj</snm><fnm>S</fnm></au>
	<au><snm>Richardson</snm><fnm>LJ</fnm></au>
	<au><snm>Finn</snm><fnm>RD</fnm></au>
	<au><snm>Bateman</snm><fnm>A</fnm></au>
	</aug>
	<pubdate>2021</pubdate>
	<volume>49</volume>
	<fpage>D412</fpage>
	<lpage>-D419</lpage>
	<url>https://academic.oup.com/nar/article/49/D1/D412/5943818</url>
	<note>Publisher: Oxford Academic</note>
	</bibl>
	
	<bibl id="B2">
	<title><p>Rfam 14: expanded coverage of metagenomic, viral and {microRNA}
	families</p></title>
	<aug>
	<au><snm>Kalvari</snm><fnm>I</fnm></au>
	<au><snm>Nawrocki</snm><fnm>EP</fnm></au>
	<au><snm>Ontiveros Palacios</snm><fnm>N</fnm></au>
	<au><snm>Argasinska</snm><fnm>J</fnm></au>
	<au><snm>Lamkiewicz</snm><fnm>K</fnm></au>
	<au><snm>Marz</snm><fnm>M</fnm></au>
	<au><snm>Griffiths Jones</snm><fnm>S</fnm></au>
	<au><snm>Toffano Nioche</snm><fnm>C</fnm></au>
	<au><snm>Gautheret</snm><fnm>D</fnm></au>
	<au><snm>Weinberg</snm><fnm>Z</fnm></au>
	<au><snm>Rivas</snm><fnm>E</fnm></au>
	<au><snm>Eddy</snm><fnm>SR</fnm></au>
	<au><snm>Finn</snm><fnm>R</fnm></au>
	<au><snm>Bateman</snm><fnm>A</fnm></au>
	<au><snm>Petrov</snm><fnm>AI</fnm></au>
	</aug>
	<pubdate>2021</pubdate>
	<volume>49</volume>
	<fpage>D192</fpage>
	<lpage>-D200</lpage>
	<url>https://doi.org/10.1093/nar/gkaa1047</url>
	</bibl>
	
	<bibl id="B3">
	<title><p>Highly accurate protein structure prediction with
	AlphaFold</p></title>
	<aug>
	<au><snm>Jumper</snm><fnm>J</fnm></au>
	<au><snm>Evans</snm><fnm>R</fnm></au>
	<au><snm>Pritzel</snm><fnm>A</fnm></au>
	<au><snm>Green</snm><fnm>T</fnm></au>
	<au><snm>Figurnov</snm><fnm>M</fnm></au>
	<au><snm>Ronneberger</snm><fnm>O</fnm></au>
	<au><snm>Tunyasuvunakool</snm><fnm>K</fnm></au>
	<au><snm>Bates</snm><fnm>R</fnm></au>
	<au><snm>{\v{Z}}{\'\i}dek</snm><fnm>A</fnm></au>
	<au><snm>Potapenko</snm><fnm>A</fnm></au>
	<au><cnm>others</cnm></au>
	</aug>
	<source>Nature</source>
	<publisher>Nature Publishing Group</publisher>
	<pubdate>2021</pubdate>
	<fpage>1</fpage>
	<lpage>-11</lpage>
	</bibl>
	
	<bibl id="B4">
	<title><p>Accurate prediction of protein structures and interactions using a
	three-track neural network</p></title>
	<aug>
	<au><snm>Baek</snm><fnm>M</fnm></au>
	<au><snm>DiMaio</snm><fnm>F</fnm></au>
	<au><snm>Anishchenko</snm><fnm>I</fnm></au>
	<au><snm>Dauparas</snm><fnm>J</fnm></au>
	<au><snm>Ovchinnikov</snm><fnm>S</fnm></au>
	<au><snm>Lee</snm><fnm>GR</fnm></au>
	<au><snm>Wang</snm><fnm>J</fnm></au>
	<au><snm>Cong</snm><fnm>Q</fnm></au>
	<au><snm>Kinch</snm><fnm>LN</fnm></au>
	<au><snm>Schaeffer</snm><fnm>RD</fnm></au>
	<au><snm>Mill{\'a}n</snm><fnm>C</fnm></au>
	<au><snm>Park</snm><fnm>H</fnm></au>
	<au><snm>Adams</snm><fnm>C</fnm></au>
	<au><snm>Glassman</snm><fnm>CR</fnm></au>
	<au><snm>DeGiovanni</snm><fnm>A</fnm></au>
	<au><snm>Pereira</snm><fnm>JH</fnm></au>
	<au><snm>Rodrigues</snm><fnm>AV</fnm></au>
	<au><snm>Dijk</snm><fnm>AA</fnm></au>
	<au><snm>Ebrecht</snm><fnm>AC</fnm></au>
	<au><snm>Opperman</snm><fnm>DJ</fnm></au>
	<au><snm>Sagmeister</snm><fnm>T</fnm></au>
	<au><snm>Buhlheller</snm><fnm>C</fnm></au>
	<au><snm>Pavkov Keller</snm><fnm>T</fnm></au>
	<au><snm>Rathinaswamy</snm><fnm>MK</fnm></au>
	<au><snm>Dalwadi</snm><fnm>U</fnm></au>
	<au><snm>Yip</snm><fnm>CK</fnm></au>
	<au><snm>Burke</snm><fnm>JE</fnm></au>
	<au><snm>Garcia</snm><fnm>KC</fnm></au>
	<au><snm>Grishin</snm><fnm>NV</fnm></au>
	<au><snm>Adams</snm><fnm>PD</fnm></au>
	<au><snm>Read</snm><fnm>RJ</fnm></au>
	<au><snm>Baker</snm><fnm>D</fnm></au>
	</aug>
	<source>Science</source>
	<publisher>American Association for the Advancement of Science</publisher>
	<pubdate>2021</pubdate>
	<volume>eabj8754</volume>
	</bibl>
	
	<bibl id="B5">
	<title><p>Identification of direct residue contacts in
	protein{\textendash}protein interaction by message passing</p></title>
	<aug>
	<au><snm>Weigt</snm><fnm>M</fnm></au>
	<au><snm>White</snm><fnm>RA</fnm></au>
	<au><snm>Szurmant</snm><fnm>H</fnm></au>
	<au><snm>Hoch</snm><fnm>JA</fnm></au>
	<au><snm>Hwa</snm><fnm>T</fnm></au>
	</aug>
	<source>Proceedings of the National Academy of Sciences</source>
	<publisher>National Academy of Sciences</publisher>
	<pubdate>2009</pubdate>
	<volume>106</volume>
	<issue>1</issue>
	<fpage>67</fpage>
	<lpage>-72</lpage>
	<url>https://www.pnas.org/content/106/1/67</url>
	</bibl>
	
	<bibl id="B6">
	<title><p>Direct-coupling analysis of residue coevolution captures native
	contacts across many protein families</p></title>
	<aug>
	<au><snm>Morcos</snm><fnm>F</fnm></au>
	<au><snm>Pagnani</snm><fnm>A</fnm></au>
	<au><snm>Lunt</snm><fnm>B</fnm></au>
	<au><snm>Bertolino</snm><fnm>A</fnm></au>
	<au><snm>Marks</snm><fnm>DS</fnm></au>
	<au><snm>Sander</snm><fnm>C</fnm></au>
	<au><snm>Zecchina</snm><fnm>R</fnm></au>
	<au><snm>Onuchic</snm><fnm>JN</fnm></au>
	<au><snm>Hwa</snm><fnm>T</fnm></au>
	<au><snm>Weigt</snm><fnm>M</fnm></au>
	</aug>
	<source>Proceedings of the National Academy of Sciences</source>
	<publisher>National Acad Sciences</publisher>
	<pubdate>2011</pubdate>
	<volume>108</volume>
	<issue>49</issue>
	<fpage>E1293</fpage>
	<lpage>-E1301</lpage>
	</bibl>
	
	<bibl id="B7">
	<title><p>Inverse statistical physics of protein sequences: a key issues
	review</p></title>
	<aug>
	<au><snm>Cocco</snm><fnm>S</fnm></au>
	<au><snm>Feinauer</snm><fnm>C</fnm></au>
	<au><snm>Figliuzzi</snm><fnm>M</fnm></au>
	<au><snm>Monasson</snm><fnm>R</fnm></au>
	<au><snm>Weigt</snm><fnm>M</fnm></au>
	</aug>
	<pubdate>2018</pubdate>
	<volume>81</volume>
	<issue>3</issue>
	<fpage>032601</fpage>
	<url>https://doi.org/10.1088/1361-6633/aa9965</url>
	<note>Publisher: {IOP} Publishing</note>
	</bibl>
	
	<bibl id="B8">
	<title><p>From residue coevolution to protein conformational ensembles and
	functional dynamics</p></title>
	<aug>
	<au><snm>Sutto</snm><fnm>L</fnm></au>
	<au><snm>Marsili</snm><fnm>S</fnm></au>
	<au><snm>Valencia</snm><fnm>A</fnm></au>
	<au><snm>Gervasio</snm><fnm>FL</fnm></au>
	</aug>
	<pubdate>2015</pubdate>
	<volume>112</volume>
	<issue>44</issue>
	<fpage>13567</fpage>
	<lpage>-13572</lpage>
	<url>http://www.pnas.org/lookup/doi/10.1073/pnas.1508584112</url>
	</bibl>
	
	<bibl id="B9">
	<title><p>Structural propensities of kinase family proteins from a Potts
	model of residue co-variation</p></title>
	<aug>
	<au><snm>Haldane</snm><fnm>A</fnm></au>
	<au><snm>Flynn</snm><fnm>WF</fnm></au>
	<au><snm>He</snm><fnm>P</fnm></au>
	<au><snm>Vijayan</snm><fnm>R. S. K.</fnm></au>
	<au><snm>Levy</snm><fnm>RM</fnm></au>
	</aug>
	<pubdate>2016</pubdate>
	<volume>25</volume>
	<issue>8</issue>
	<fpage>1378</fpage>
	<lpage>-1384</lpage>
	<url>https://onlinelibrary.wiley.com/doi/abs/10.1002/pro.2954</url>
	</bibl>
	
	<bibl id="B10">
	<title><p>How Pairwise Coevolutionary Models Capture the Collective Residue
	Variability in Proteins?</p></title>
	<aug>
	<au><snm>Figliuzzi</snm><fnm>M</fnm></au>
	<au><snm>Barrat Charlaix</snm><fnm>P</fnm></au>
	<au><snm>Weigt</snm><fnm>M</fnm></au>
	</aug>
	<pubdate>2018</pubdate>
	<volume>35</volume>
	<issue>4</issue>
	<fpage>1018</fpage>
	<lpage>-1027</lpage>
	<url>https://doi.org/10.1093/molbev/msy007</url>
	</bibl>
	
	<bibl id="B11">
	<title><p>Sparse generative modeling via parameter reduction of Boltzmann
	machines: Application to protein-sequence families</p></title>
	<aug>
	<au><snm>Barrat Charlaix</snm><fnm>P</fnm></au>
	<au><snm>Muntoni</snm><fnm>AP</fnm></au>
	<au><snm>Shimagaki</snm><fnm>K</fnm></au>
	<au><snm>Weigt</snm><fnm>M</fnm></au>
	<au><snm>Zamponi</snm><fnm>F</fnm></au>
	</aug>
	<source>Physical Review E</source>
	<publisher>APS</publisher>
	<pubdate>2021</pubdate>
	<volume>104</volume>
	<issue>2</issue>
	<fpage>024407</fpage>
	</bibl>
	
	<bibl id="B12">
	<title><p>Distance-based protein folding powered by deep learning</p></title>
	<aug>
	<au><snm>Xu</snm><fnm>J</fnm></au>
	</aug>
	<source>Proceedings of the National Academy of Sciences</source>
	<publisher>National Acad Sciences</publisher>
	<pubdate>2019</pubdate>
	<volume>116</volume>
	<issue>34</issue>
	<fpage>16856</fpage>
	<lpage>-16865</lpage>
	</bibl>
	
	<bibl id="B13">
	<title><p>Deep learning extends de novo protein modelling coverage of genomes
	using iteratively predicted structural constraints</p></title>
	<aug>
	<au><snm>Greener</snm><fnm>JG</fnm></au>
	<au><snm>Kandathil</snm><fnm>SM</fnm></au>
	<au><snm>Jones</snm><fnm>DT</fnm></au>
	</aug>
	<source>Nature communications</source>
	<publisher>Nature Publishing Group</publisher>
	<pubdate>2019</pubdate>
	<volume>10</volume>
	<issue>1</issue>
	<fpage>1</fpage>
	<lpage>-13</lpage>
	</bibl>
	
	<bibl id="B14">
	<title><p>Improved protein structure prediction using potentials from deep
	learning</p></title>
	<aug>
	<au><snm>Senior</snm><fnm>AW</fnm></au>
	<au><snm>Evans</snm><fnm>R</fnm></au>
	<au><snm>Jumper</snm><fnm>J</fnm></au>
	<au><snm>Kirkpatrick</snm><fnm>J</fnm></au>
	<au><snm>Sifre</snm><fnm>L</fnm></au>
	<au><snm>Green</snm><fnm>T</fnm></au>
	<au><snm>Qin</snm><fnm>C</fnm></au>
	<au><snm>{\v{Z}}{\'\i}dek</snm><fnm>A</fnm></au>
	<au><snm>Nelson</snm><fnm>AW</fnm></au>
	<au><snm>Bridgland</snm><fnm>A</fnm></au>
	<au><cnm>others</cnm></au>
	</aug>
	<source>Nature</source>
	<publisher>Nature Publishing Group</publisher>
	<pubdate>2020</pubdate>
	<volume>577</volume>
	<issue>7792</issue>
	<fpage>706</fpage>
	<lpage>-710</lpage>
	</bibl>
	
	<bibl id="B15">
	<title><p>Improved protein structure prediction using predicted interresidue
	orientations</p></title>
	<aug>
	<au><snm>Yang</snm><fnm>J</fnm></au>
	<au><snm>Anishchenko</snm><fnm>I</fnm></au>
	<au><snm>Park</snm><fnm>H</fnm></au>
	<au><snm>Peng</snm><fnm>Z</fnm></au>
	<au><snm>Ovchinnikov</snm><fnm>S</fnm></au>
	<au><snm>Baker</snm><fnm>D</fnm></au>
	</aug>
	<source>Proceedings of the National Academy of Sciences</source>
	<publisher>National Acad Sciences</publisher>
	<pubdate>2020</pubdate>
	<volume>117</volume>
	<issue>3</issue>
	<fpage>1496</fpage>
	<lpage>-1503</lpage>
	</bibl>
	
	<bibl id="B16">
	<title><p>An evolution-based model for designing chorismate mutase
	enzymes</p></title>
	<aug>
	<au><snm>Russ</snm><fnm>WP</fnm></au>
	<au><snm>Figliuzzi</snm><fnm>M</fnm></au>
	<au><snm>Stocker</snm><fnm>C</fnm></au>
	<au><snm>Barrat Charlaix</snm><fnm>P</fnm></au>
	<au><snm>Socolich</snm><fnm>M</fnm></au>
	<au><snm>Kast</snm><fnm>P</fnm></au>
	<au><snm>Hilvert</snm><fnm>D</fnm></au>
	<au><snm>Monasson</snm><fnm>R</fnm></au>
	<au><snm>Cocco</snm><fnm>S</fnm></au>
	<au><snm>Weigt</snm><fnm>M</fnm></au>
	<au><snm>Ranganathan</snm><fnm>R</fnm></au>
	</aug>
	<pubdate>2020</pubdate>
	<volume>369</volume>
	<issue>6502</issue>
	<fpage>440</fpage>
	<lpage>-445</lpage>
	<url>https://science.sciencemag.org/content/369/6502/440</url>
	<note>Publisher: American Association for the Advancement of Science Section:
	Report</note>
	</bibl>
	
	<bibl id="B17">
	<title><p>Coevolutionary landscape inference and the context-dependence of
	mutations in beta-lactamase TEM-1</p></title>
	<aug>
	<au><snm>Figliuzzi</snm><fnm>M</fnm></au>
	<au><snm>Jacquier</snm><fnm>H</fnm></au>
	<au><snm>Schug</snm><fnm>A</fnm></au>
	<au><snm>Tenaillon</snm><fnm>O</fnm></au>
	<au><snm>Weigt</snm><fnm>M</fnm></au>
	</aug>
	<source>Molecular biology and evolution</source>
	<publisher>Oxford University Press</publisher>
	<pubdate>2016</pubdate>
	<volume>33</volume>
	<issue>1</issue>
	<fpage>268</fpage>
	<lpage>-280</lpage>
	</bibl>
	
	<bibl id="B18">
	<title><p>Mutation effects predicted from sequence co-variation</p></title>
	<aug>
	<au><snm>Hopf</snm><fnm>TA</fnm></au>
	<au><snm>Ingraham</snm><fnm>JB</fnm></au>
	<au><snm>Poelwijk</snm><fnm>FJ</fnm></au>
	<au><snm>Sch{\"a}rfe</snm><fnm>CP</fnm></au>
	<au><snm>Springer</snm><fnm>M</fnm></au>
	<au><snm>Sander</snm><fnm>C</fnm></au>
	<au><snm>Marks</snm><fnm>DS</fnm></au>
	</aug>
	<source>Nature biotechnology</source>
	<publisher>Nature Publishing Group</publisher>
	<pubdate>2017</pubdate>
	<volume>35</volume>
	<issue>2</issue>
	<fpage>128</fpage>
	</bibl>
	
	<bibl id="B19">
	<title><p>A learning algorithm for Boltzmann machines</p></title>
	<aug>
	<au><snm>Ackley</snm><fnm>DH</fnm></au>
	<au><snm>Hinton</snm><fnm>GE</fnm></au>
	<au><snm>Sejnowski</snm><fnm>TJ</fnm></au>
	</aug>
	<source>Cognitive science</source>
	<publisher>Elsevier</publisher>
	<pubdate>1985</pubdate>
	<volume>9</volume>
	<issue>1</issue>
	<fpage>147</fpage>
	<lpage>-169</lpage>
	</bibl>
	
	<bibl id="B20">
	<title><p>Training products of experts by minimizing contrastive
	divergence</p></title>
	<aug>
	<au><snm>Hinton</snm><fnm>GE</fnm></au>
	</aug>
	<source>Neural computation</source>
	<publisher>MIT Press</publisher>
	<pubdate>2002</pubdate>
	<volume>14</volume>
	<issue>8</issue>
	<fpage>1771</fpage>
	<lpage>-1800</lpage>
	</bibl>
	
	<bibl id="B21">
	<title><p>Information Theory and Statistical Mechanics</p></title>
	<aug>
	<au><snm>Jaynes</snm><fnm>E. T.</fnm></au>
	</aug>
	<source>Phys. Rev.</source>
	<pubdate>1957</pubdate>
	<volume>106</volume>
	<issue>4</issue>
	<fpage>620</fpage>
	<lpage>-630</lpage>
	<url>https://link.aps.org/doi/10.1103/PhysRev.106.620</url>
	<note>Publisher: American Physical Society</note>
	</bibl>
	
	<bibl id="B22">
	<title><p>Correlation-compressed direct-coupling analysis</p></title>
	<aug>
	<au><snm>Gao</snm><fnm>CY</fnm></au>
	<au><snm>Zhou</snm><fnm>HJ</fnm></au>
	<au><snm>Aurell</snm><fnm>E</fnm></au>
	</aug>
	<pubdate>2018</pubdate>
	<volume>98</volume>
	<issue>3</issue>
	<fpage>032407</fpage>
	<url>https://link.aps.org/doi/10.1103/PhysRevE.98.032407</url>
	<note>Publisher: American Physical Society</note>
	</bibl>
	
	<bibl id="B23">
	<title><p>Direct-coupling analysis of residue coevolution captures native
	contacts across many protein families</p></title>
	<aug>
	<au><snm>Morcos</snm><fnm>F</fnm></au>
	<au><snm>Pagnani</snm><fnm>A</fnm></au>
	<au><snm>Lunt</snm><fnm>B</fnm></au>
	<au><snm>Bertolino</snm><fnm>A</fnm></au>
	<au><snm>Marks</snm><fnm>DS</fnm></au>
	<au><snm>Sander</snm><fnm>C</fnm></au>
	<au><snm>Zecchina</snm><fnm>R</fnm></au>
	<au><snm>Onuchic</snm><fnm>JN</fnm></au>
	<au><snm>Hwa</snm><fnm>T</fnm></au>
	<au><snm>Weigt</snm><fnm>M</fnm></au>
	</aug>
	<pubdate>2011</pubdate>
	<volume>108</volume>
	<issue>49</issue>
	<fpage>E1293</fpage>
	<lpage>-E1301</lpage>
	<url>https://www.pnas.org/content/108/49/E1293</url>
	<note>Publisher: National Academy of Sciences Section: {PNAS} Plus</note>
	</bibl>
	
	<bibl id="B24">
	<title><p>Equation of State Calculations by Fast Computing
	Machines</p></title>
	<aug>
	<au><snm>Metropolis</snm><fnm>N</fnm></au>
	<au><snm>Rosenbluth</snm><fnm>AW</fnm></au>
	<au><snm>Rosenbluth</snm><fnm>MN</fnm></au>
	<au><snm>Teller</snm><fnm>AH</fnm></au>
	<au><snm>Teller</snm><fnm>E</fnm></au>
	</aug>
	<source>J. Chem. Phys.</source>
	<pubdate>1953</pubdate>
	<volume>21</volume>
	<issue>6</issue>
	<fpage>1087</fpage>
	<lpage>-1092</lpage>
	<url>https://aip.scitation.org/doi/10.1063/1.1699114</url>
	<note>Publisher: American Institute of Physics</note>
	</bibl>
	
	<bibl id="B25">
	<title><p>Monte Carlo sampling methods using Markov chains and their
	applications</p></title>
	<aug>
	<au><snm>Hastings</snm><fnm>W. K.</fnm></au>
	</aug>
	<source>Biometrika</source>
	<pubdate>1970</pubdate>
	<volume>57</volume>
	<issue>1</issue>
	<fpage>97</fpage>
	<lpage>-109</lpage>
	<url>https://doi.org/10.1093/biomet/57.1.97</url>
	</bibl>
	
	<bibl id="B26">
	<title><p>Stochastic relaxation, {G}ibbs distributions, and the {B}ayesian
	restoration of images</p></title>
	<aug>
	<au><snm>Geman</snm><fnm>S.</fnm></au>
	<au><snm>Geman</snm><fnm>D.</fnm></au>
	</aug>
	<source>{IEEE} Trans Pattern Anal Mach Intell</source>
	<pubdate>1984</pubdate>
	<volume>6</volume>
	<issue>6</issue>
	<fpage>721</fpage>
	<lpage>-741</lpage>
	</bibl>
	
	<bibl id="B27">
	<title><p>Adaptive subgradient methods for online learning and stochastic
	optimization</p></title>
	<aug>
	<au><snm>Duchi</snm><fnm>J</fnm></au>
	<au><snm>Hazan</snm><fnm>E</fnm></au>
	<au><snm>Singer</snm><fnm>Y</fnm></au>
	</aug>
	<source>COLT 2010 - The 23rd Conference on Learning Theory</source>
	<series><title><p>COLT 2010 - The 23rd Conference on Learning
	Theory</p></title></series>
	<pubdate>2010</pubdate>
	<fpage>257</fpage>
	<lpage>-269</lpage>
	</bibl>
	
	<bibl id="B28">
	<title><p>Note on learning rate schedules for stochastic
	optimization</p></title>
	<aug>
	<au><snm>Darken</snm><fnm>C</fnm></au>
	<au><snm>Moody</snm><fnm>J</fnm></au>
	</aug>
	<source>Proceedings of the 3rd International Conference on Neural Information
	Processing Systems</source>
	<publisher>Morgan Kaufmann Publishers Inc.</publisher>
	<series><title><p>{NIPS}'90</p></title></series>
	<pubdate>1990</pubdate>
	<fpage>832</fpage>
	<lpage>-838</lpage>
	</bibl>
	
	<bibl id="B29">
	<title><p>Translating {HIV} Sequences into Quantitative Fitness Landscapes
	Predicts Viral Vulnerabilities for Rational Immunogen Design</p></title>
	<aug>
	<au><snm>Ferguson</snm><fnm>AL</fnm></au>
	<au><snm>Mann</snm><fnm>JK</fnm></au>
	<au><snm>Omarjee</snm><fnm>S</fnm></au>
	<au><snm>Ndung’u</snm><fnm>T</fnm></au>
	<au><snm>Walker</snm><fnm>BD</fnm></au>
	<au><snm>Chakraborty</snm><fnm>AK</fnm></au>
	</aug>
	<source>Immunity</source>
	<pubdate>2013</pubdate>
	<volume>38</volume>
	<issue>3</issue>
	<fpage>606</fpage>
	<lpage>-617</lpage>
	<url>https://www.sciencedirect.com/science/article/pii/S1074761313001076</url>
	</bibl>
	
	<bibl id="B30">
	<title><p>Structural propensities of kinase family proteins from a Potts
	model of residue co-variation</p></title>
	<aug>
	<au><snm>Haldane</snm><fnm>A.</fnm></au>
	<au><snm>Flynn</snm><fnm>W.F.</fnm></au>
	<au><snm>He</snm><fnm>P.</fnm></au>
	<au><snm>Vijayan</snm><fnm>R.S.K.</fnm></au>
	<au><snm>Levy</snm><fnm>R.M.</fnm></au>
	</aug>
	<source>Protein Science</source>
	<pubdate>2016</pubdate>
	<fpage>1378</fpage>
	<lpage>-1384</lpage>
	</bibl>
	
	<bibl id="B31">
	<title><p>Mi3-{GPU}: {MCMC}-based inverse Ising inference on {GPUs} for
	protein covariation analysis</p></title>
	<aug>
	<au><snm>Haldane</snm><fnm>A</fnm></au>
	<au><snm>Levy</snm><fnm>RM</fnm></au>
	</aug>
	<source>Computer Physics Communications</source>
	<pubdate>2021</pubdate>
	<volume>260</volume>
	<fpage>107312</fpage>
	<url>https://linkinghub.elsevier.com/retrieve/pii/S0010465520301193</url>
	</bibl>
	
	<bibl id="B32">
	<title><p>Structural Relaxation Made Simple</p></title>
	<aug>
	<au><snm>Bitzek</snm><fnm>E</fnm></au>
	<au><snm>Koskinen</snm><fnm>P</fnm></au>
	<au><snm>Gähler</snm><fnm>F</fnm></au>
	<au><snm>Moseler</snm><fnm>M</fnm></au>
	<au><snm>Gumbsch</snm><fnm>P</fnm></au>
	</aug>
	<source>Phys. Rev. Lett.</source>
	<pubdate>2006</pubdate>
	<volume>97</volume>
	<issue>17</issue>
	<fpage>170201</fpage>
	<url>https://link.aps.org/doi/10.1103/PhysRevLett.97.170201</url>
	</bibl>
	
	<bibl id="B33">
	<title><p>Improved contact prediction in proteins: Using pseudolikelihoods to
	infer {P}otts models</p></title>
	<aug>
	<au><snm>Ekeberg</snm><fnm>M</fnm></au>
	<au><snm>Lövkvist</snm><fnm>C</fnm></au>
	<au><snm>Lan</snm><fnm>Y</fnm></au>
	<au><snm>Weigt</snm><fnm>M</fnm></au>
	<au><snm>Aurell</snm><fnm>E</fnm></au>
	</aug>
	<source>Phys. Rev. E</source>
	<pubdate>2013</pubdate>
	<volume>87</volume>
	<issue>1</issue>
	<fpage>012707</fpage>
	<url>https://link.aps.org/doi/10.1103/PhysRevE.87.012707</url>
	<note>Publisher: American Physical Society</note>
	</bibl>
	
	<bibl id="B34">
	<title><p>Assessing the accuracy of direct-coupling analysis for {RNA}
	contact prediction</p></title>
	<aug>
	<au><snm>Cuturello</snm><fnm>F</fnm></au>
	<au><snm>Tiana</snm><fnm>G</fnm></au>
	<au><snm>Bussi</snm><fnm>G</fnm></au>
	</aug>
	<source>{RNA}</source>
	<pubdate>2020</pubdate>
	<fpage>rna.074179.119</fpage>
	<url>http://rnajournal.cshlp.org/content/early/2020/02/27/rna.074179.119</url>
	</bibl>
	
	<bibl id="B35">
	<title><p>infernet-h2020/pfam\_interactions: Initial release</p></title>
	<aug>
	<au><snm>Sarti</snm><fnm>E</fnm></au>
	<au><snm>Pagnani</snm><fnm>A</fnm></au>
	</aug>
	<publisher>Zenodo</publisher>
	<pubdate>2020</pubdate>
	</bibl>
	
	<bibl id="B36">
	<source>\url{https://www.ncbi.nlm.nih.gov/}</source>
	</bibl>
	
	<bibl id="B37">
	<title><p>Equilibrium and non-Equilibrium regimes in the learning of
	Restricted Boltzmann Machines</p></title>
	<aug>
	<au><snm>Decelle</snm><fnm>A</fnm></au>
	<au><snm>Furtlehner</snm><fnm>C</fnm></au>
	<au><snm>Seoane</snm><fnm>B</fnm></au>
	</aug>
	<pubdate>2021</pubdate>
	<url>http://arxiv.org/abs/2105.13889</url>
	</bibl>
	
	</refgrp>
} 

\end{document}